\documentclass[12pt]{article}
\usepackage{amssymb}
\usepackage{graphicx}
\usepackage[cp1251]{inputenc}
\usepackage{rotating}

\begin{document}
 \noindent {\footnotesize\it Astronomy Letters, 2023, Vol. 49, No. 7, pp. 410--420}
 \newcommand{\dif}{\textrm{d}}

 \noindent
 \begin{tabular}{llllllllllllllllllllllllllllllllllllllllllllll}
 & & & & & & & & & & & & & & & & & & & & & & & & & & & & & & & & & & & & & &\\\hline\hline
 \end{tabular}

  \vskip 0.5cm
  \bigskip
 \bigskip
\centerline{\large\bf 
 Peculiarities of Open Star Clusters with High Vertical Velocities}
\centerline{\large\bf from the Region of the Sco-Cen OB Association}

 \bigskip
 \bigskip
  \centerline { %  DOI: 10.1134/S1063773723070010
   V. V. Bobylev\footnote [1]{bob-v-vzz@rambler.ru},  A. T. Bajkova}
 \bigskip
 \centerline{\small\it Pulkovo Astronomical Observatory, Russian Academy of Sciences, St. Petersburg, 196140 Russia}
 \bigskip
 \bigskip
{We have studied the kinematics of a unique sample of young open star clusters (OSCs) with high vertical velocities, $15<W<40$~km s$^{-1}$. The characteristics of these clusters were taken from the catalogue by Hunt and Reffert (2023), where their mean proper motions, line-of-sight velocities, and distances were calculated using Gaia~DR3 data. These OSCs are located within 0.6~kpc of the Sun and form two clumps: one in the region of the Sco-Cen OB association and the other one in the region
of the Per~OB3--Per~OB2 associations. The OSC group of 47 members in the region of the Sco-Cen association is shown to expand along the $y$ axis, $\partial V/\partial y=51\pm12$~km s$^{-1}$ kpc$^{-1}$. This group also has a positive rotation around the $z$ axis with an angular velocity of $71\pm11$~km s$^{-1}$ kpc$^{-1}$ and a negative rotation around the $x$ axis with an angular velocity of $-35\pm5$~km s$^{-1}$ kpc$^{-1}$. Based on the velocities of 27 OSCs from the region of the Per~OB3--Per~OB2 associations, we have found no gradients differing significantly from zero. We have studied the kinematics of more than 1700 stars selected by Luhman (2022) as probable members of the Sco-Cen OB association. These stars are shown to have no high vertical velocities. The expansion coefficient of the stellar system in the $xy$ plane has been found from all stars to be $K_{xy}=43.2\pm2.2$~km s$^{-1}$ kpc$^{-1}$. Based on stars from the three UCL, LCC, and V1062 Sco groups with a mean age $\sim$20 Myr, for the first time we have found a volume expansion coefficient of the stellar system
differing significantly from zero, $K_{xyz}=43.2\pm3.4$~km s$^{-1}$ kpc$^{-1}$.
 }

\bigskip
\section*{INTRODUCTION}
Studying the internal structure of OB associations is of great importance for solving various astrophysical problems. For example, these include
(a) estimating the formation rate of clusters of different
ages, (b) refining the details of the expansion
and dispersal of associations during the formation
of a population of field stars, (c) the possibility
of the formation of rich associations and clusters
through the merging of smaller structures, (d) determining
the structure of clusters and the duration of
their formation in star-forming regions, etc. A huge
number of publications are devoted to the investigation
of OB associations (see, e.g., Blaauw 1954;
de Zeeuw et al. 1999; Preibish and Zinnecker 1999; Zinnecker and Yorke 2007; Feigelson 2017; Mel'nik and Dambis 2020; Wright 2020; Dobbs et al. 2022).

The expansion effect, which is one of the important kinematic properties, has been recorded in many OB associations (Cantat-Gaudin et al. 2019; Rao et al. 2020; Armstrong et al. 2020, 2022). The bulk rotation of an OB association is harder to estimate. For example, Kuhn et al. (2019) studied a sample
of 28 clusters and associations with ages from 1 to 5 Myr, the data for which were taken from the Gaia DR2 catalogue (Brown et al. 2018). These authors
showed that at least 75\% of these systems expand
with typical velocities of 0.5~km s$^{-1}$, while bulk
rotation was detected only in one system. According to the estimates by Mel'nik and Dambis (2017, 2018), the PerOB1 and CarOB1 OB associations expand with linear velocities $\sim$6~km s$^{-1}$ at their outer boundaries.

The OB associations have a complex structure. They contain gas, dust, stars, and open star clusters (OSCs). Studying OSCs is of great importance for
solving various problems, in particular, for understanding
the structure and evolution of associations.
At present, one of the most extensive collections of
kinematic data on Galactic OSCs is the catalogue
by Hunt and Reffert (2023). It presents 7200 OSCs
that were identified from Gaia~DR3 data (Vallenari et al. 2022) using the popular HDBSCAN cluster analysis algorithm. As it turned out, in this catalogue 4780 OSCs are already known from the literature,
while 2420 from the total number of those found are
new OSC candidates. The ages, lifetimes, and distances
were estimated for all OSCs in the catalogue.
A large percentage of OSCs have an estimate of
theirmean line-of-sight velocity from GaiaDR3 data.
In Bobylev and Bajkova (2023) OSCs younger than
50 Myr were taken from the catalogue by Hunt and
Reffert (2023) to analyze the kinematics of the Galaxy
and the spiral density wave. A total of 2494 OSCs
with a mean age of 21Myr were selected; 1722 OSCs
of them have line-of-sight velocities. A big surprise
was the detection of a strong spike in the vertical
velocities of OSCs in a relatively small solar neighborhood.
This effect can be clearly seen in Fig. 1. The periodic perturbations in the vertical OSC velocities $f_W$ of this OSC sample associated with the Galactic
spiral density wave are also present but have a very small amplitude, $f_W=1.1\pm0.4$~km s$^{-1}$. The periodic perturbations found are indicated by the red line in Fig.~1.

The goal of this paper is to study the spatial distribution and kinematics of young OSCs close to the Sun with high vertical velocities $W.$ For this
purpose, we use an approach based on the linear
Ogorodnikov-Milne model. We also apply this approach
to analyze the stars belonging to the Sco-Cen
OB association closest to the Sun.
The paper is structured as follows. In the next
``Method'' Section we describe our approach based
on the linear Ogorodnikov-Milne model. Next, in
the ``Results'' Section we first describe the results of
our analysis of the young OSCs with high vertical
velocities selected by us. We took the characteristics
of these OSCs the catalogue by Hunt and Reffert
(2023). Then, we present the results of our analysis
of the probable member stars of the Sco-Cen
OB association taken from the list by Luhman (2022).
In the ``Conclusions'' we formulate our main results.

%%%%%%%%%%%%%%%%%%%%%%%%%%%%%%%%%%%%%%%%%%%%%%%%%%%%%%%%%%%%% F1:
\begin{figure}[t]
{ \begin{center}
  \includegraphics[width=0.6\textwidth]{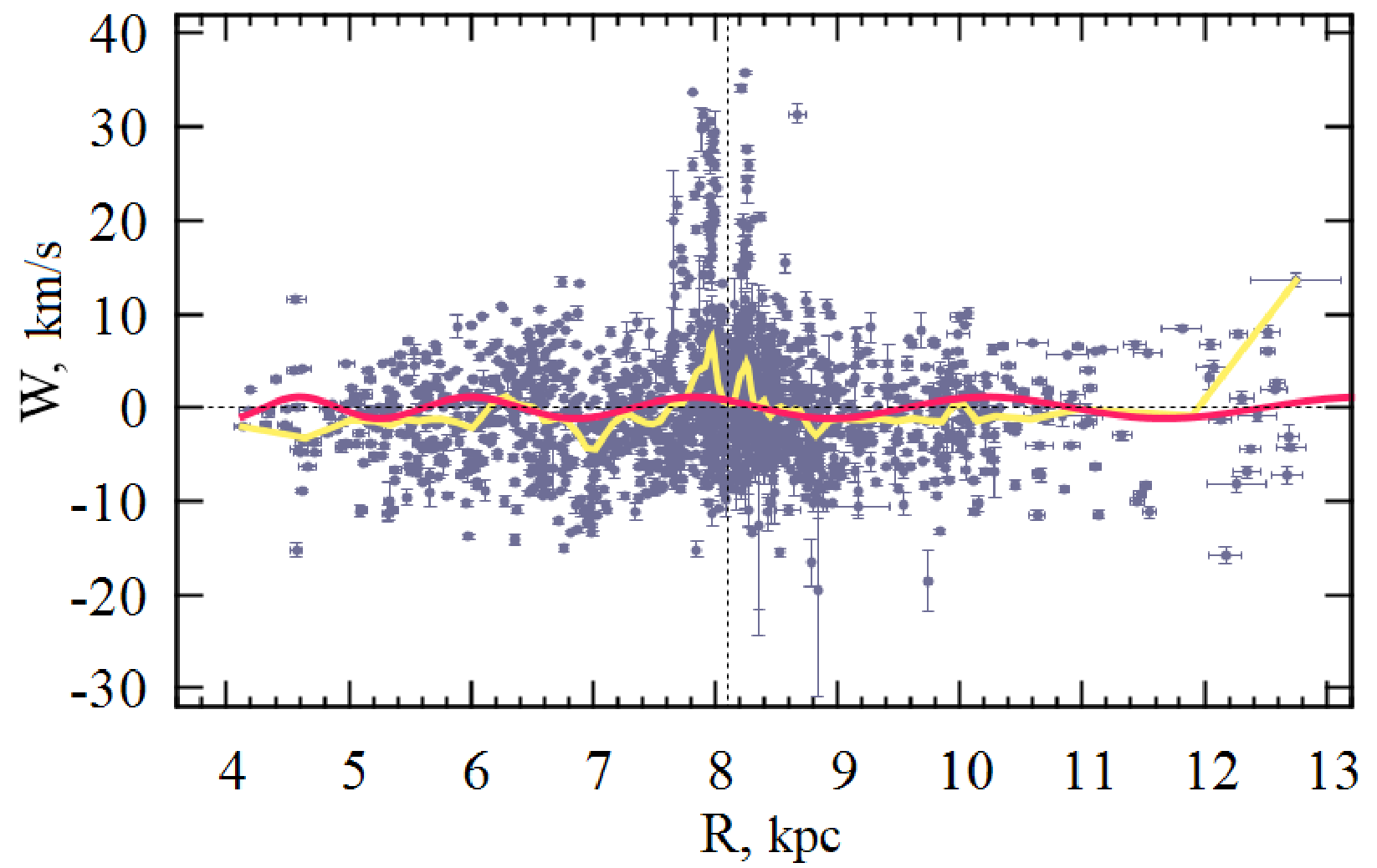}
  \caption{
Vertical velocity $W$ versus distance $R$ for the sample of OSCs younger than 50~Myr. The averaged data are represented by the yellow line; the periodic curve found from our spectral analysis is indicated by the red line. The figure was taken from Bobylev and Bajkova (2023).}
 \label{WR-50}
\end{center}}
\end{figure}
%%%%%%%%%%%%%%%%%%%%  f1

\section*{THE METHOD}
We have three stellar velocity components from observations: the line-of-sight velocity $V_r$ and the two tangential velocity components $V_l=4.74r\mu_l\cos b$ and $V_b=4.74r\mu_b$ along the Galactic longitude $l$ and latitude $b,$ respectively, expressed in km s$^{-1}$. Here, the coefficient 4.74 is the ratio of the number of kilometers in an astronomical unit to the number of seconds in a
tropical year, and r is the stellar heliocentric distance
in kpc. The proper motion components $\mu_l\cos b$ and $\mu_b$
are expressed in mas yr$^{-1}$ (milliarcseconds per year).
The velocities $U,V,W$ directed along the rectangular
Galactic coordinate axes $x,y,z$ are calculated via
the components $V_r, V_l, V_b$:
 \begin{equation}
 \begin{array}{lll}
 U=V_r\cos l\cos b-V_l\sin l-V_b\cos l\sin b,\\
 V=V_r\sin l\cos b+V_l\cos l-V_b\sin l\sin b,\\
 W=V_r\sin b                +V_b\cos b,
 \label{UVW}
 \end{array}
 \end{equation}
where the velocity $U$ and the coordinate $x$ are directed from the Sun toward the Galactic center, $V$ and $y$ are in the direction of Galactic rotation, and W and z are directed to the north Galactic pole.

According to the linear Ogorodnikov-Milne model (Ogorodnikov 1965), assuming that the peculiar velocity of the Sun $(U,V,W)_\odot$ is eliminated or zero, the velocities $U,V,W$ can be represented as the following system of linear equations:
 \begin{equation}
 \renewcommand{\arraystretch}{2.6}
 \begin{array}{lll}
 \displaystyle
 \qquad
 U={\left(\frac{\partial U}{\partial x}\right)}_0  x+
   {\left(\frac{\partial U}{\partial y}\right)}_0  y+
   {\left(\frac{\partial U}{\partial z}\right)}_0  z, \\
 \displaystyle
 \qquad
 V={\left(\frac{\partial V}{\partial x}\right)}_0  x+
   {\left(\frac{\partial V}{\partial y}\right)}_0  y+
   {\left(\frac{\partial V}{\partial z}\right)}_0  z, \\
 \displaystyle
 \qquad
 W={\left(\frac{\partial W}{\partial x}\right)}_0  x+
   {\left(\frac{\partial W}{\partial y}\right)}_0  y+
   {\left(\frac{\partial W}{\partial z}\right)}_0  z.
 \label{model-1}
 \end{array}
 \end{equation}
Here, the subscript ``0'' means that the derivatives are
taken at the coordinate origin. Based on the three
diagonal gradients from the right-hand sides of the
system of equations (2), we can find the expansion
parameters of the stellar system $K_{xy}$, $K_{xz}$ and $K_{yz}$
in the three corresponding planes:
 \begin{equation}
 \renewcommand{\arraystretch}{2.6}
 \begin{array}{lll}
 \displaystyle
 \qquad
  K_{xy}=\left[
   {\left(\frac{\partial U}{\partial x}\right)}_0+
   {\left(\frac{\partial V}{\partial y}\right)}_0 \right]/2, \\
\displaystyle
 \qquad
  K_{xz}=\left[
   {\left(\frac{\partial U}{\partial x}\right)}_0+
   {\left(\frac{\partial W}{\partial z}\right)}_0 \right]/2, \\
\displaystyle
 \qquad
  K_{yz}=\left[
   {\left(\frac{\partial Y}{\partial x}\right)}_0+
   {\left(\frac{\partial W}{\partial z}\right)}_0 \right]/2,
 \label{Kxy}
 \end{array}
 \end{equation}
and the volume expansion coefficient
 \begin{equation}
 \renewcommand{\arraystretch}{2.2}
 \begin{array}{lll}
 \displaystyle
 \qquad
 K_{xyz}=\left[
   {\left(\frac{\partial U}{\partial x}\right)}_0+
   {\left(\frac{\partial V}{\partial y}\right)}_0+
   {\left(\frac{\partial W}{\partial z}\right)}_0 \right]/3.
 \label{K-xyz}
 \end{array}
 \end{equation}
Note that all nine gradients involved in the system of
equations (2) can be found graphically.

%%%%%%%%%%%%%%%%%%%%%%%%%%%%%%%%%%%%%%%%%%%%%%%%%%%%%%%%%%%%% F2:
\begin{figure}[t]
{ \begin{center}
  \includegraphics[width=0.6\textwidth]{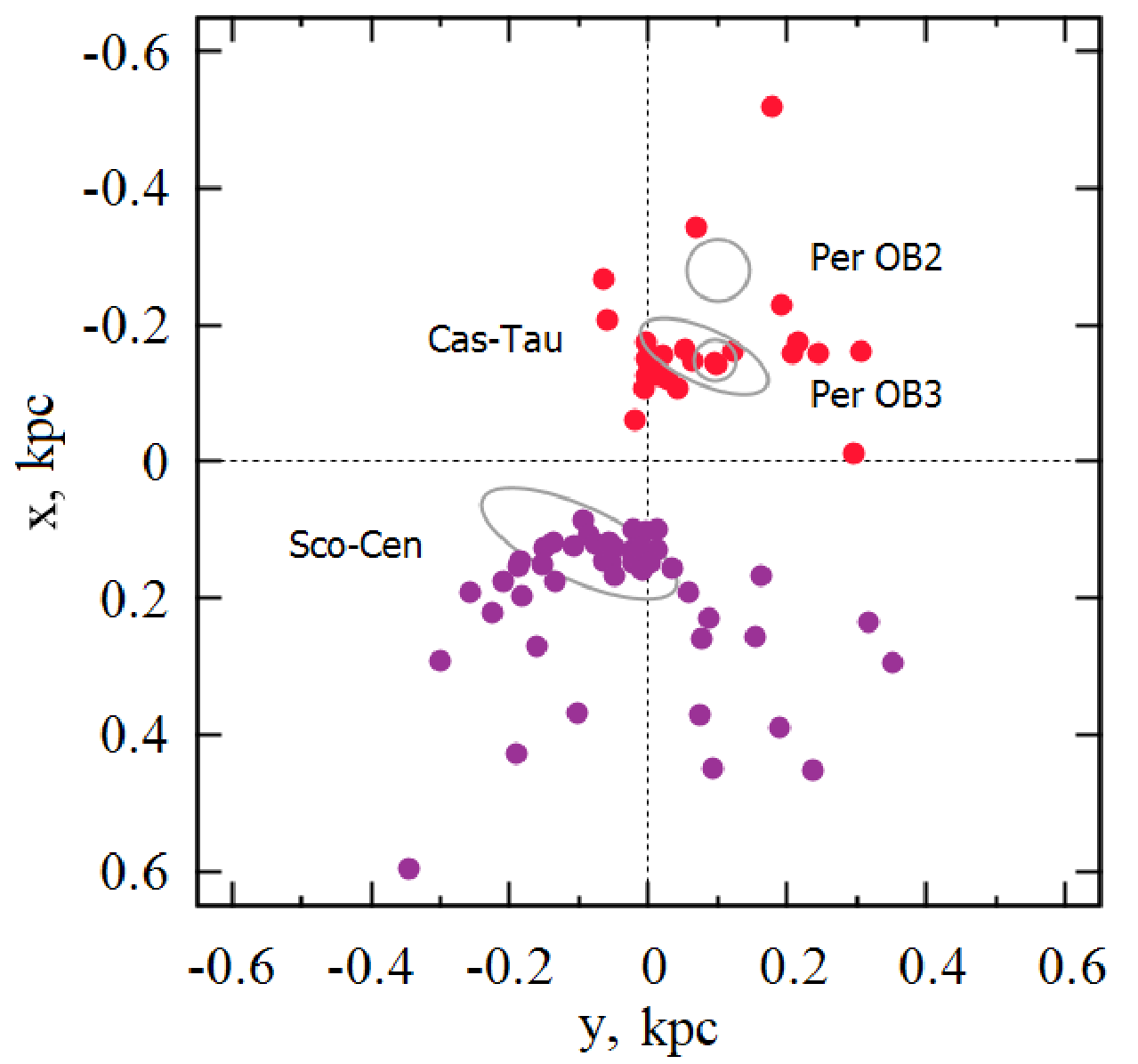}
  \caption{
Distribution of OSCs with high ($W>15$~km s$^{-1}$) vertical velocities in projection onto the Galactic $xy$ plane; the gray ellipses indicate approximate positions of the Sco-Cen, PerOB3, Cas-Tau, and PerOB2 OB associations.
}
 \label{XY}
\end{center}}
\end{figure}
%%%%%%%%%%%%%%%%%%%%  f2
%%%%%%%%%%%%%%%%%%%%%%%%%%%%%%%%%%%%%%%%%%%%%%%%%%%%%%%%%%%%% F3:
\begin{figure}[t]
{ \begin{center}
 \includegraphics[width=0.85\textwidth]{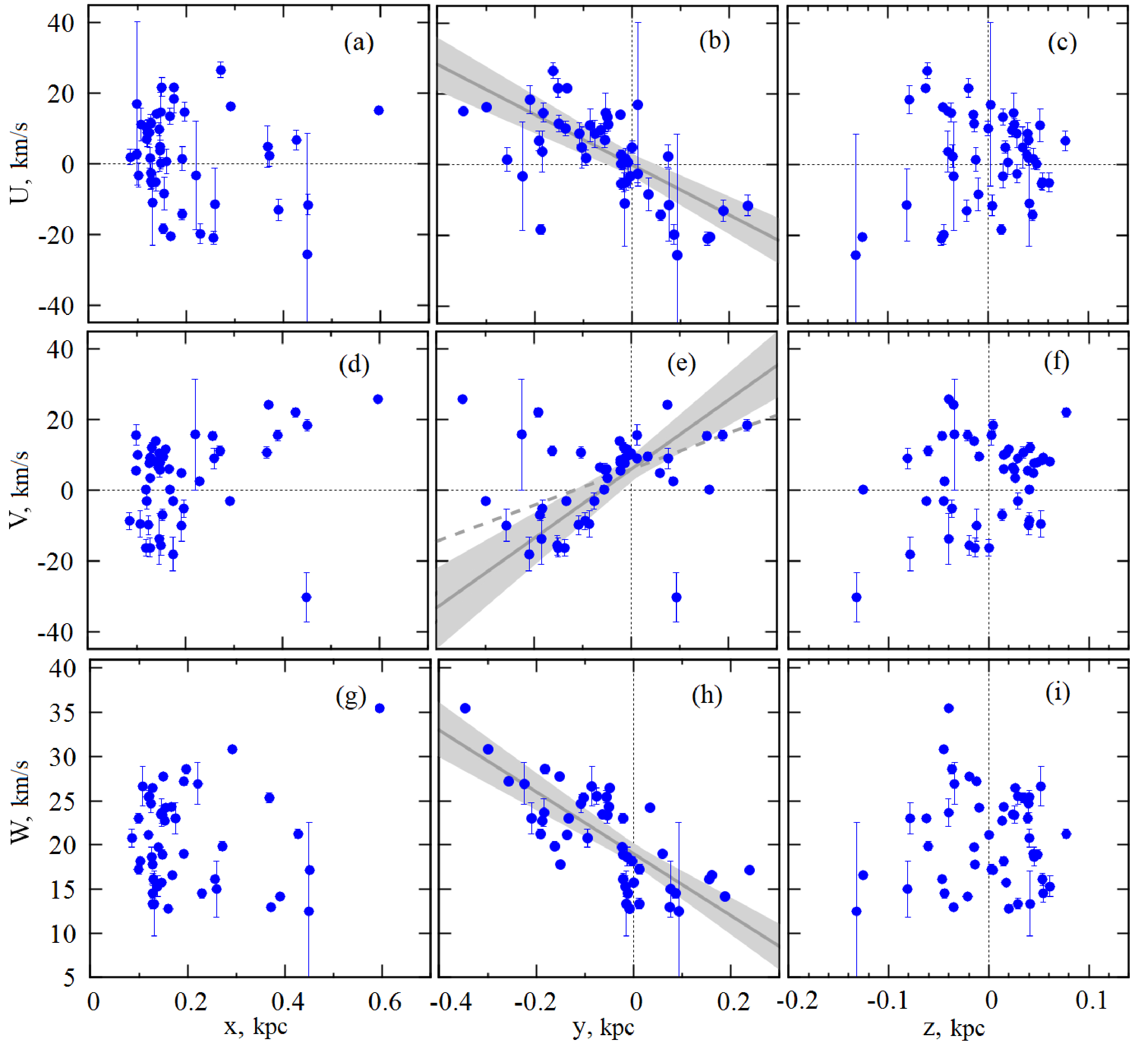}
\caption{Velocities $U,V,W$ versus coordinates $x,y,z$ for the selected OSCs with high ($W>15$~km s$^{-1}$) vertical velocities from the region of the Sco-Cen association.}
 \label{XYZ-UVW-SCO}
\end{center}}
\end{figure}
%%%%%%%%%%%%%%%%%%%%  f3
%%%%%%%%%%%%%%%%%%%%%%%%%%%%%%%%%%%%%%%%%%%%%%%%%%%%%%%%%%%%% F4:
\begin{figure}[t]
{ \begin{center}
  \includegraphics[width=0.85\textwidth]{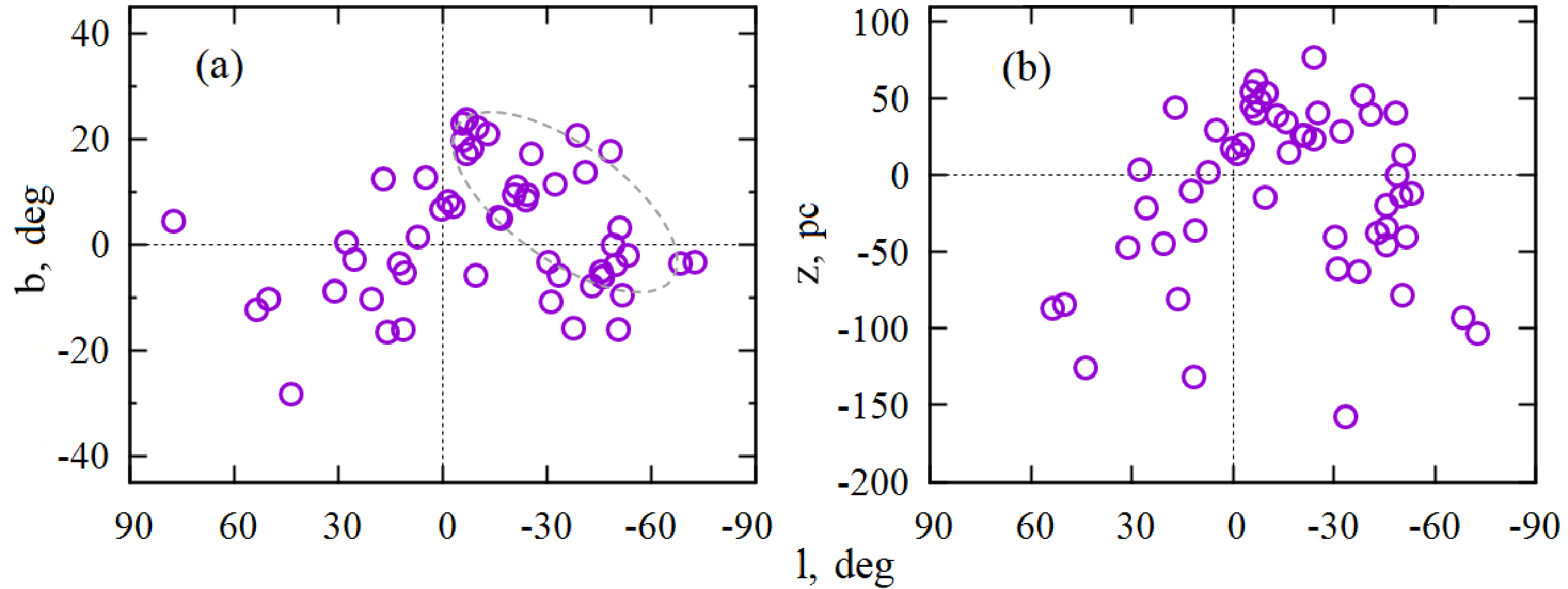}
\caption{(a) Distribution of the selected OSCs near the region of the Sco-Cen association in the $lb$ plane, the ellipse indicates
approximate positions of the stars in this association. (b) The distribution of these OSCs in the $lz$ plane.
}
 \label{lb-Sco}
\end{center}}
\end{figure}
%%%%%%%%%%%%%%%%%%%%  f4
%%%%%%%%%%%%%%%%%%%%%%%%%%%%%%%%%%%%%%%%%%%%%%%%%%%%%%%%%%%%% F5:
\begin{figure}[t]
{ \begin{center}
  \includegraphics[width=0.85\textwidth]{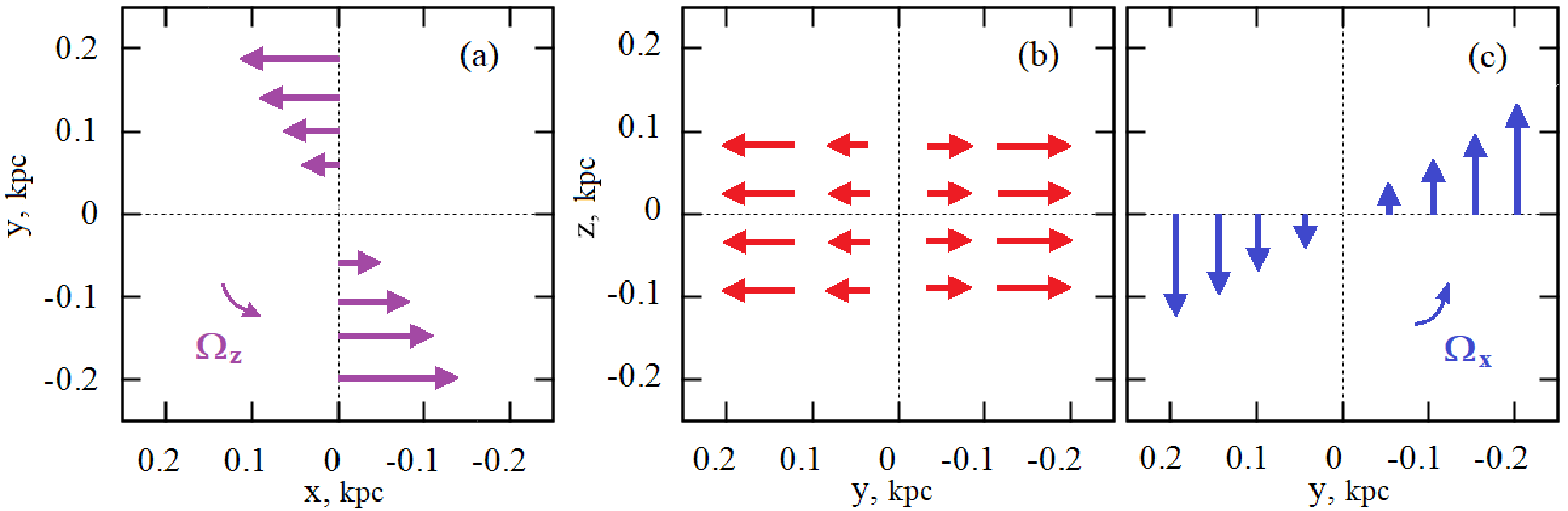}
\caption{Schematic distribution of the velocities $U,V,W$ for OSCs from the region of the Sco-Cen association: (a) the velocity $U$ in the $xy$ plane, (b) the velocity $V$ in the $yz$ plane, and (c) the velocity $W$ in the $yz$ plane.
}
 \label{U-Sco}
\end{center}}
\end{figure}
%%%%%%%%%%%%%%%%%%%%  f5
%%%%%%%%%%%%%%%%%%%%%%%%%%%%%%%%%%%%%%%%%%%%%%%%%%%%%%%%%%%%% F6:
\begin{figure}[t]
{ \begin{center}
  \includegraphics[width=0.85\textwidth]{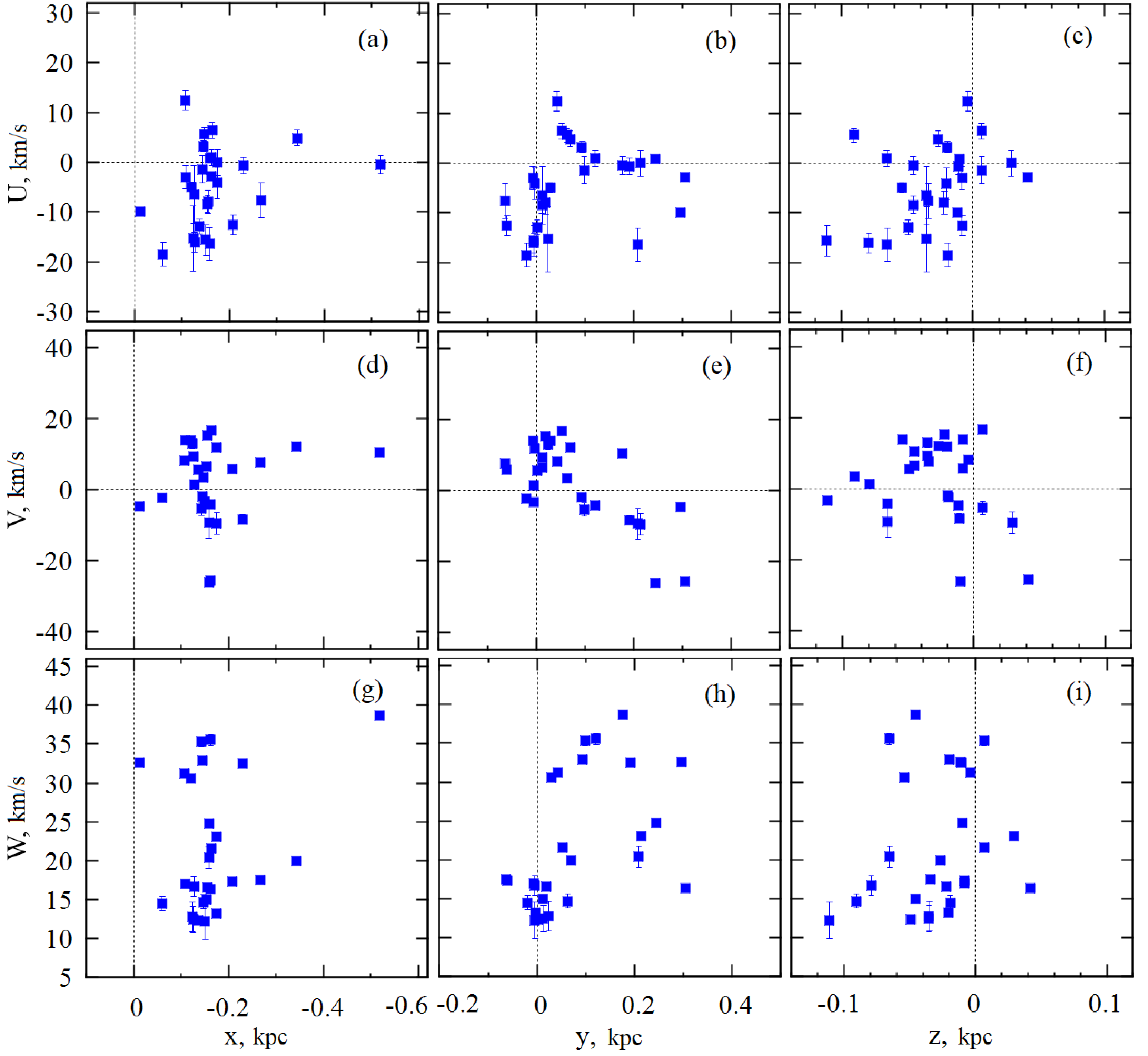}
\caption{Velocities $U,V,W$ versus coordinates $x,y,z$ for the selected OSCs with high vertical velocities from the region of the
Per\,OB3--Per\,OB2 associations.
}
 \label{XYZ-UVW-PER}
\end{center}}
\end{figure}
%%%%%%%%%%%%%%%%%%%%  f6
%%%%%%%%%%%%%%%%%%%%%%%%%%%%%%%%%%%%%%%%%%%%%%%%%%%%%%%%%%%%% F7:
\begin{figure}[t]
{ \begin{center}
  \includegraphics[width=0.85\textwidth]{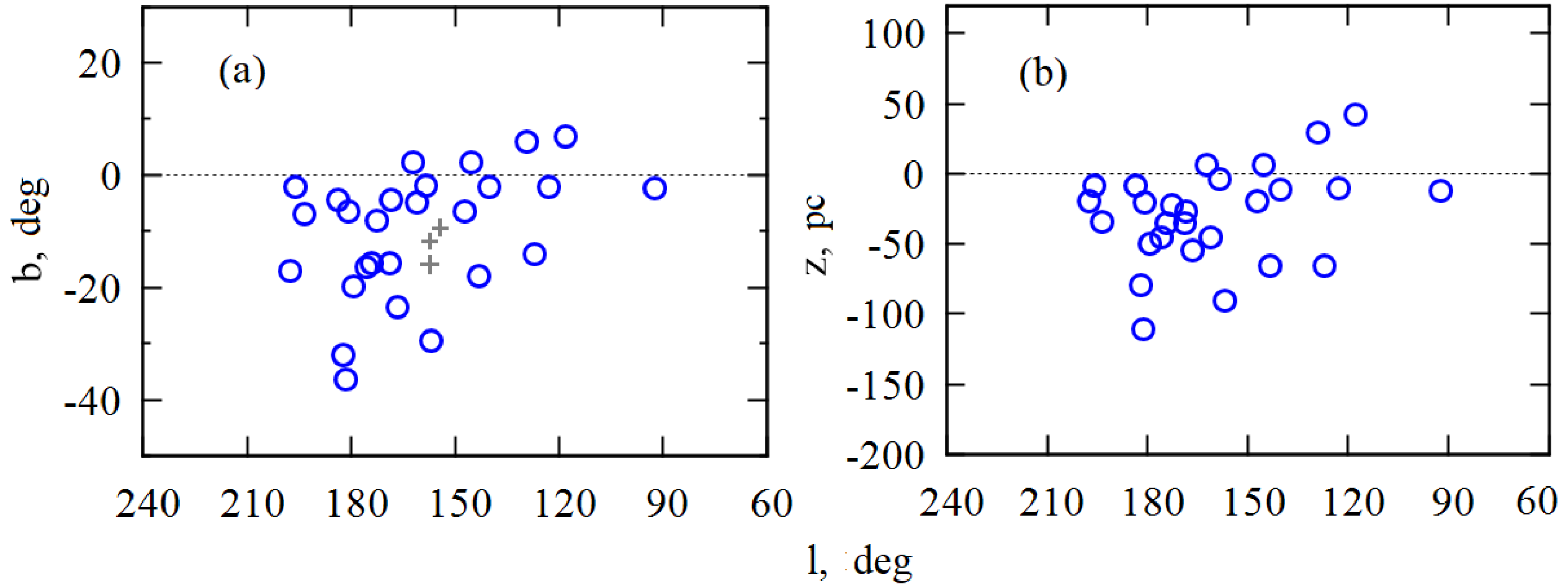}
\caption{(a) Distribution of the selected OSCs near the Per\,OB3--Per\,OB2 region in the $lb$ plane, the crosses mark the centers of the Per OB3, Cas-Tau, and PerOB2 associations. (b) The distribution of these OSCs in the $lz$ plane.
}
 \label{lb-PER}
\end{center}}
\end{figure}
%%%%%%%%%%%%%%%%%%%%  f7

\section*{RESULTS}
At the beginning, using Fig. 1 as a guide, we selected OSCs with high vertical velocities by applying the constraint $W>15$~km s$^{-1}$; we also used
the constraint on the OSC age $<80$~Myr. As it
turned out, all of the selected OSCs lie either in the
Scorpio-Centaurus region or in the Taurus-Perseus
region. We designate the group of OSCs as Sco-Cen
and PerOB3-PerOB2 in the first and second cases,
respectively. Figure 2 presents the distribution of
the OSCs selected in this way in projection onto the
Galactic $xy$ plane. The ellipses indicate approximate
positions of the Sco-Cen, PerOB3, Cas-Tau, and
PerOB2 OB associations. When specifying the positions
of the associations in space and on the celestial sphere, we are oriented to the papers by de Zeeuw et al. (1999) and Mel'nik and Dambis (2020).

The Sco-Cen OB association is usually divided into three parts: US (Upper Scorpius), UCL (Upper Centaurus Lupus), and LCC (Lower Centaurus Crux), whose centers are at distances of 145, 140, and 118 pc from the Sun, respectively. Almost all of the
stars in the Sco-Cen OB association are located in the fourth Galactic quadrant.

In the PerOB3--PerOB2 region under consideration there are three OB associations whose centers lie virtually on one line of sight: the compact PerOB3
($\alpha$Per) association, the considerably larger (in occupied
space) Cas-Tau association, and the PerOB2
association. According to de Zeeuw et al. (1999),
the motion of the $\alpha$Per cluster is consistent with the
motion of Cas-Tau, suggesting a physical connection
between these two stellar groups. Thus, the halo
of the PerOB3 association is the inner region of the
Cas-Tau association.

The PerOB3 association is located at a distance of about 180 pc from the Sun toward $l\sim150^\circ$ and $b\sim-7^\circ$. The distance to the center of the Cas-Tau
association is about 200 pc, the center lies toward $l\sim160^\circ$ and $b\sim-10^\circ$. The farthest of the described three associations, Per OB2, is at a distance of about 300 pc, its center lies toward $l\sim160^\circ$ and $b\sim-16^\circ$.

\subsection*{The Region of the Sco--Cen Association}
In Fig. 3 the velocities $U,VW$ are plotted against the coordinates $x,y,z$ for the selected OSCs with high vertical velocities from the region of the Sco-Cen association.

The group in the region of the Sco-Cen association includes the following 47 OSCs: HSC 151, HSC 157, HSC 191, HSC 199, HSC 310, HSC 376, HSC 2585, HSC 2603, HSC 2615, HSC 2618, HSC 2630, HSC 2636, HSC 2648, HSC 2662,
HSC 2690, HSC 2733, HSC 2782, HSC 2816,
HSC 2850, HSC 2900, HSC 2907, HSC 2919,
HSC 2931, HSC 2963, HSC 2976, Harvard 10,
OC0666, OCSN92, OCSN96, OCSN 98, OCSN 100,
Platais 10, Theia 67, Theia 222, Theia 436, BH 164,
CWNU 1004, CWNU 1143, HSC 67, HSC 103,
Theia 711, Theia 828, UPK 33, UPK 606, UPK 612,
UPK 624, and UPK 640 with a mean age of 28~Myr.
The distribution of these OSCs on the celestial sphere
is presented in Fig.~4. As follows from this figure as well as from Fig.~2, most of the OSCs are located near the place occupied by the Sco-Cen association.

From the velocities of these 47 OSCs from the region of the Sco-Cen association we found three gradients differing significantly from zero:
 $\partial U/\partial y=-71\pm11$~km s$^{-1}$ kpc$^{-1}$, 
 $\partial V/\partial y= 51\pm12$~km s$^{-1}$ kpc$^{-1}$, and 
 $\partial W/\partial y=-35\pm5$~km s$^{-1}$ kpc$^{-1}$. These gradients were found by solving conditional equations of the form $y=a+bx$
(here, $y$ and $x$ are the corresponding velocities and
coordinates, $a$ is the constant term, and $b$ is the corresponding
gradient) by the least-squares method with two unknowns, $a$ and $b.$ The dashed line in Fig.~3e indicates the dependence with a slope $\partial V/\partial y=98\pm20$~km s$^{-1}$ kpc$^{-1}$ derived from 30 OSCs under the constraint $|y|<0.15$~kpc, i.e., from the core of the spatial cluster distribution.

Out of the three diagonal terms, only one gradient is nonzero, $\partial V/\partial y=51\pm12$~km s$^{-1}$ kpc$^{-1}$. This suggests that the subsystem of OSCs being analyzed expands along the $y$ axis.

The gradients $\partial U/\partial y$ and $\partial W/\partial y$ are related
to the rotation of the stellar system. According to the linear Ogorodnikov-Milne model (Ogorodnikov 1965), in the absence of deformations, the
components of the angular velocity $\Omega_{x,y,z}$ around the
corresponding coordinate axes can be found from the
system of conditional equations (2):
 \begin{equation}
 \renewcommand{\arraystretch}{1.4}
 \begin{array}{lll}
  U=\quad~~~~~   0-\Omega_z\cdot y+\Omega_y\cdot z,\\
  V=~~~\Omega_z\cdot x+ ~~~~~   0-\Omega_x\cdot z,\\
  W=-\Omega_y\cdot x+\Omega_x\cdot y+ ~~~~~    0.
 \label{model-Omega}
 \end{array}
 \end{equation}
As a result, we have a positive rotation around the $z$ axis with an angular velocity $\Omega_z=-\partial U/\partial y=71\pm11$~km s$^{-1}$ kpc$^{-1}$ and a negative rotation around the $x$ axis with an angular velocity
 $\Omega_x=\partial W/\partial y=-35\pm5$~km s$^{-1}$ kpc$^{-1}$.

Figure 5 presents a schematic distribution of the velocities $U,$ $V,$ and $W$ for OSCs from the region of the Sco-Cen association in the three corresponding
planes. In the $xy$ plane the distribution of the vectors qualitatively reflects the gradient $\partial U/\partial y$ found, the thin arrow indicates the rotation $\Omega_z$ in a direction opposite to the Galactic rotation. The $V$ distribution
in Fig.~5b reflects the expansion of the OSC system along the $y$ axis in accordance with the gradient $\partial V/\partial y$.

The $W$ distribution in Fig.~5c qualitatively reflects
the gradient $\partial W/\partial y$ found, the thin arrow indicates
the direction of rotation around the $x$ axis, $\Omega_x$. Of
course, it should be kept in mind that here all of the
OSCs being analyzed have high vertical velocities,
while the rotation occurs relative to their arbitrary
center with a velocity $W\approx20$~km s$^{-1}$.

\subsection*{The Region of the PerOB3--PerOB2 Association}
In Fig. 6 the velocities $U,V,W$ are plotted against
the coordinates $x,y,z$ for the selected OSCs with
high vertical velocities from the PerOB3--PerOB2
region. From the velocities of the OSCs from this
region we found no gradients differing significantly
from zero.

This group includes the following 27 OSCs:
CWNU 1129, HSC 940, HSC 981, HSC 1020,
HSC 1040, HSC 1131, HSC 1165, HSC 1238,
HSC 1254, HSC 1270, HSC 1279, HSC 1314,
HSC 1318, HSC 1426, HSC 1438, HSC 1531,
HSC 1553, HSC 1566, Melotte 20, Melotte 22,
Theia 7, Theia 54, Theia 65, Theia 66, Theia 93, Theia 517, and UPK 303 with a mean age of 33 Myr. The distribution of theseOSCs on the celestial sphere
is presented in Fig.~7. Note that Melotte 20 is the well-known $\alpha$Per OSC, while Melotte 22 is the Pleiades.

\subsection*{Stars from the Sco--Cen Association}
It is interesting to compare the kinematic characteristics
of OSCs with relatively high vertical velocities
in the region of the Sco-Cen OB association derived
in this paper with the kinematic characteristics
of member stars of this association. It is important
to find out how close the kinematic characteristics of
the OSCs being analyzed are to the characteristics
of the stars that are the most probable members of
the association. After all, in the case of a weak
and unrestricted correlation of such characteristics,
mechanisms that are barely related to the processes
describing the internal kinematics of the associations
should be invoked to understand the detected properties
of the OSC sample.

The selection of stars belonging to the Sco-Cen OB association has been made repeatedly on the basis of various observational data (Blaauw 1954;
de Zeeuw et al. 1999; Sartori et al. 2003; Mel'nik and
Dambis 2017; Damiani et al. 2019; Luhman 2022).

The paper by Luhman (2022), where faint stars that are probable members of the Sco-Cen OB association were selected, is among the last ones in this list. The selection was made using the spatial positions of the stars, kinematic, and photometric criteria. The catalogue contains 10\,509 stars with
Gaia EDR3 data, more than 1700 of these stars have
line-of-sight velocities whose values were taken by
Luhman from the literature. The stars with distance,
proper motion, and line-of-sight velocity estimates
are of greatest interest to us, since they allow a full-fledged
three-dimensional analysis of their kinematics to be performed. Note that the presence of gradients $\partial U/\partial x$ and $\partial V/\partial y$ can be seen in Fig.~16 from Luhman (2022).

As members of the Sco-Cen OB association Luhman considered the following six subsystems: UCL, LCC, US, Lupus, V1062 Sco, and Ophiuchus. These subsystems have slightly different mean ages: UCL--16 Myr (Pecaut and Mamajek 2016), LCC--15 Myr (Pecaut and Mamajek 2016), US--10 Myr (Luhman and Esplin 2020), Lupus--6 Myr (Luhman
2022), V1062 Sco--20 Myr (Luhman 2022), and Ophiuchus--2-6 Myr (Luhman 2022). It is
important to note that, according to Luhman (2022), the age of UCL and LCC is $\sim$20~Myr. It is also well known that the age estimates from low-mass
stars differ from those from high-mass stars. For example, the age of US is $\sim$5 Myr from low-mass stars and $\sim$11 Myr from intermediate/high-mass
stars (Herczeg and Hillenbrand 2015).

We see that the Sco-Cen OB association includes three very young ($\sim$6~Myr) subsystems--US, Lupus, and Ophiuchus (which are located nearby on the celestial sphere as well), and three older ($\sim$20~Myr) subsystems--UCL, LCC, and V1062 Sco.

In Figs. 8a--8c the velocities $U,$ $V,$ and $W$ are plotted against the coordinates $x,$ $y,$ and $z,$ respectively, for the complete sample of stars from the Sco-Cen association, with the stars from the US, Lupus,
and Ophiuchus subsystems being highlighted
by the pink color. From all stars of this sample we
found the following two gradients differing significantly
from zero: $\partial U/\partial x=31.3\pm2.6$~km s$^{-1}$ kpc$^{-1}$
and $\partial V/\partial y=55.2\pm1.6$~km s$^{-1}$ kpc$^{-1}$. These dependences
are presented on the corresponding panels with the specified $1\sigma$ confidence region (here, $\sigma$ is the root-mean-square deviation from the linear dependence found). We used the boundary value $y>-0.03$~kpc as the only condition for the selection of stars belonging to the US, Lupus, and Ophiuchus subsystems. In this case, we have an expansion coefficient of the stellar system in the $xy$ plane (3) differing significantly from zero:
\begin{equation}
K_{xy}=43.2\pm2.2~\hbox{km s$^{-1}$ kpc$^{-1}$}.
 \label{K-x-y}
\end{equation}
In Figs. 8d--8f the velocities $U,$ $V,$ and $W$ are plotted against the coordinates $x,$ $y,$ and $z,$ respectively, for the sample of stars from the Sco-Cen association that does not include the US, Lupus, and Ophiuchus members. From the stars
of this sample we found the following three gradients:
 $\partial U/\partial x=44.6\pm3.3$~km s$^{-1}$ kpc$^{-1}$, 
 $\partial V/\partial y=44.9\pm3.8$~km s$^{-1}$ kpc$^{-1}$, and 
 $\partial W/\partial z=40.2\pm3.1$~km s$^{-1}$ kpc$^{-1}$. These dependences are presented on the corresponding panels with the specified $1\sigma$
confidence region (shading). In this case, we have not
only an expansion coefficient differing significantly
from zero in each of the three planes (3), but also a
volume expansion coefficient of the stellar system (4):
\begin{equation}
K_{xyz}=43.2\pm3.4~\hbox{km s$^{-1}$ kpc$^{-1}$}.
 \label{K-x-y-z}
\end{equation}

 \section{DISCUSSION}
In their kinematic analysis of stars in the Sco-Cen association Bobylev and Bajkova (2020) found the expansion coefficient of the system to be $K=39\pm2$~km s$^{-1}$ kpc$^{-1}$. This coefficient has the meaning of Kxy (see Eq. (3)), since the approach with four Oort constants ($A, B, C,$ and $K$) that describe the kinematics in this plane was used.
 
In contrast, the determination of the three-dimensional
expansion coefficient $K_{xyz}$ (see Eq. (4)) can
be encountered very rarely in the literature. In this
connection, it is interesting to note the paper by Armstrong
et al. (2020), where the following three gradients
were found for population B of young stars from
the region of the VelOB2 association: 
 $\partial U/\partial x=98\pm21$~km s$^{-1}$ kpc$^{-1}$, 
 $\partial V/\partial y=44\pm7$~km s$^{-1}$ kpc$^{-1}$, and 
 $\partial W/\partial z=69\pm11$~km s$^{-1}$ kpc$^{-1}$. In that case,
we can estimate $K_{xyz}=70\pm8$~km s$^{-1}$ kpc$^{-1}$. Taking
the radius of the association to be 0.05 kpc, according
to Fig.~7 from the paper of these authors,
we estimate the linear volume expansion velocity of
the association at its outer boundary to be $3.5\pm0.4$~km s$^{-1}$. 
 
One of the first expansion effects of the Sco-Cen OB association was estimated by Blaauw (1964). From the data on young massive B-type stars he found the rate of its linear expansion to be $K_{xy}=50$~km s$^{-1}$ kpc$^{-1}$, which allowed the expansion age to be estimated, 20~Myr. 
 
It is interesting to note the paper by Wright
and Mamajek (2018), who tested the kinematics of
the Sco-Cen association by several methods using
data on stars from the Gaia~DR1 catalogue (Brown
et al. 2016). In particular, they considered a method
of searching for the linear expansion coefficient from
the line-of-sight velocities of stars and traced back
the orbits of stars to find the time of the smallest
region of their spatial concentration. Surprisingly,
but these authors found no evidence of expansion of
the association. 

In contrast, Goldman et al. (2018) showed an expansion of stars in the LCC subgroup with $K_{xy}\sim35$~km s$^{-1}$ kpc$^{-1}$. Bobylev and Bajkova (2007)
found $K_{xy}=46\pm8$~km s$^{-1}$ kpc$^{-1}$ from massive stars of the Sco-Cen association with data from the Hipparcos (1997) catalogue, while Bobylev and
Bajkova (2020) obtained the estimate of $K_{xy}=39\pm2$~km s$^{-1}$ kpc$^{-1}$ by taking into account the influence of the Galactic spiral density wave using stars from the Gaia~DR2 catalogue and showed the absence of a noticeable rotation of the association.

Thus, the expansion coefficient of the Sco-Cen OBassociation in the $xy$ plane found by us from stars, $K_{xy}=43.2\pm2.2$~km s$^{-1}$ kpc$^{-1}$ (6), is in good agreement with the results of other authors. In contrast,
the estimate of the volume expansion coefficient for
the older subsystem of the Sco-Cen OB association,
$K_{xyz}=43.2\pm3.4$~km s$^{-1}$ kpc$^{-1}$ (7), is a new one.

The results obtained from the analysis of stars and
OSCs in the region of the Sco-Cen OB association
show that groups differing in age have different kinematics.
The stellar core of the association consisting
of UCL, LCC, and V1062 Sco shows an almost
uniform expansion in all planes and an expansion
in all directions. The younger stellar fraction of the
association consisting of US, Lupus, and Ophiuchus
has a peculiar motion along the coordinate $z$ and
exhibits an expansion in the $xy$ plane. The oldest
part of the Sco-Cen OB association consisting of
OSCs with high vertical velocities with a mean age
$\sim$28~Myr shows an expansion along the $y$ axis and
rotations around the $x$ and $z$ axes. Based on Fig. 4b,
we get the impression that OSCs with high vertical velocities outline the outer boundary of the Sco-Cen OB association.

On the other hand, the appearance of OSCs with
high vertical velocities is probably related to a larger scale
process. As we see, a region comparable to
the region occupied by the Gould Belt ($r<0.6$~kpc)
is affected by such a process. The Radcliffe Wave
(Alves et al. 2020), which manifests itself as vertical
perturbations in the positions and velocities of young
stars and OSCs (Bobylev et al. 2022a), is also known
here, with the specific mechanism that caused such
perturbations being unknown so far. For example,
Fleck (2020) proposes to associate the origin of the
Radcliffe Wave with the Kelvin--Helmholtz instability.
Marchal and Martin (2023) consider the emergence
of the Radcliffe Wave in terms of the formation of
the North Celestial Pole Loop. However, most of
the researchers (Widrow et al. 2012; Bennett and
Bovy 2018; Thulasidharan et al. 2022) adhere to the
assumption that the perturbations in the vertical coordinates
and vertical velocities of stars in the Galactic disk could be caused by the gravitational effect of an impactor like a dwarf satellite galaxy of the Milky Way.

We found that the kinematic properties of the OSCs with high vertical velocities duffer greatly from the internal kinematics of the stars belonging to
the Sco-Cen OB association. In the next paper
we are going to select OSCs lying in the region of the Radcliffe Wave from the catalogue by Hunt and Reffert (2023) and perform their spectral analysis (as was done based on a sample of masers in Bobylev et al. (2022b)).

 \section{CONCLUSIONS}
We studied the kinematics of a unique sample of
young (younger than 80 Myr) OSCs with high vertical
velocities, $15<W<40$~km s$^{-1}$. These clusters
are located at $r<0.6$~kpc. We took the kinematic
characteristics of the OSCs from the catalogue by
Hunt and Reffert (2023), where their mean proper
motions, line-of-sight velocities, and distances were
calculated using GaiaDR3 data. We established that
the young OSCs selected in this way are located
in the region of OB associations close to the Sun.
One group is located in the region of the Sco-Cen
association, the second one is located in the region
of the Per OB3, Cas-Tau, and Per OB2 associations,
which we designated as PerOB3--PerOB2.

TheOSC group of 47members in the region of the Sco-Cen association was shown to expand along the $y$ axis, $\partial V/\partial y=51\pm12$~km s$^{-1}$ kpc$^{-1}$, with the core of the sample consisting of 30 OSCs exhibiting an
even large value, $\partial V/\partial y=98\pm20$~km s$^{-1}$ kpc$^{-1}$.
This group also has a positive rotation around the $z$ axis with an angular velocity $\Omega_z=71\pm11$~km s$^{-1}$ kpc$^{-1}$ and a negative rotation around
the $x$ axis with an angular velocity $\Omega_x=-35\pm5$~km s$^{-1}$ kpc$^{-1}$. Based on the velocities of 27 OSCs from the region of the PerOB3--PerOB2 associations, we found no gradients differing significantly from zero.

We studied $\sim$1700 stars selected by Luhman (2022) as probable members of the Sco-Cen OB association. Two gradients differing significantly from
zero were found from all stars of this sample: 
 $\partial U/\partial x=31.3\pm2.6$~km s$^{-1}$ kpc$^{-1}$ and 
$\partial V/\partial y=55.2\pm1.6$~km s$^{-1}$ kpc$^{-1}$. In this case, we have an expansion coefficient of the stellar system in the $xy$ plane differing significantly from zero: $K_{xy}=43.2\pm2.2$~km s$^{-1}$ kpc$^{-1}$. These stars have no high vertical velocities.

Based on the stars of three groups, UCL, LCC, and V1062 Sco, with a mean age $\sim$20 Myr, we found the following three gradients: 
 $\partial U/\partial x=44.6\pm3.3$~km s$^{-1}$ kpc$^{-1}$, 
 $\partial V/\partial y=44.9\pm3.8$~km s$^{-1}$ kpc$^{-1}$, and 
$\partial W/\partial z=40.2\pm3.1$~km s$^{-1}$ kpc$^{-1}$. In this case, the volume expansion coefficient of the stellar system differs significantly from zero,
 $K_{xyz}=43.2\pm3.4$~km s$^{-1}$ kpc$^{-1}$. This parameter has been estimated for the first time.

 \subsubsection*{ACKNOWLEDGMENTS}
We are grateful to the referees for their useful remarks that contributed to an improvement of the paper.

 \subsubsection*{REFERENCES}
 \small

\quad~~1. J. Alves, C. Zucker, A. A. Goodman, J. S. Speagle,
St. Meingast, Th. Robitaille, D. P. Finkbeiner, E. F. Schlafly, and G. M. Green, Nature (London, U.K.) 578, 237 (2020).

2. J. J. Armstrong, N. J. Wright, R. D. Jeffries, and R. J. Jackson, Mon. Not. R. Astron. Soc. 494, 4794 (2020).

3. J. J. Armstrong, N. J. Wright, R. D. Jeffries,
R. J. Jackson, and T. Cantat-Gaudin, Mon. Not. R. Astron. Soc. 517, 5704 (2022).

4. M. Bennett and J. Bovy, Mon. Not. R. Astron. Soc. 482, 1417 (2018).

5. A. Blaauw, Astron. J. 59, 317 (1954).

6. A. Blaauw, Ann. Rev. Astron. Astrophys. 2, 213 (1964).

7. V. V. Bobylev and A. T. Bajkova, Astron. Lett. 33, 571 (2007).

8. V. V. Bobylev and A. T. Bajkova, Astron. Rep. 64, 326 (2020).

9. V. V. Bobylev, A. T. Bajkova, and Yu. N. Mishurov, Astron. Lett. 48, 434 (2022b).

10. V. V. Bobylev, A. T. Bajkova, and Yu. N. Mishurov, Astrophysics 65, 579 (2022a).

11. V. V. Bobylev and A. T. Bajkova, Astron. Lett. 49, 320 (2023).

12. A. G. A. Brown, A. Vallenari, T. Prusti, et al. (Gaia
Collab.), Astron. Astrophys. 595, A2 (2016).

13. A. G. A. Brown, A. Vallenari, T. Prusti, et al. (Gaia
Collab.), Astron. Astrophys. 616, A1 (2018).

14. T. Cantat-Gaudin, C. Jordi, N. J. Wright, J. J. Armstrong, A. Vallenari, L. Balaguer-Nunez, P. Ramos, D. Bossini, et al., Astron. Astrophys. 626, 17 (2019).

15. F. Damiani, L. Prisinzano, I. Pillitteri, G. Micela, and
S. Sciortino, Astron. Astrophys. 623, A112 (2019).

16. C. L. Dobbs, T. J. R. Bending, A. R. Pettitt,
A. S. M. Buckner, and M. R. Bate, Mon. Not. R. Astron. Soc. 517, 675 (2022).

17. E. D. Feigelson, arXiv: 1704.0811 (2017).

18. R. Fleck, Nature (London, U.K.) 583, 24 (2020).

19. B. Goldman, S. R\"oser, E. Schilbach, A. C. Moor, and
T. Henning, Astrophys. J. 868, 32 (2018).

20. G. J. Herczeg and L. A. Hillenbrand, Astrophys. J. 808, 23 (2015).

21. The HIPPARCOS and Tycho Catalogues, ESA SP--1200 (1997).

22. E. L. Hunt and S. Reffert, Astron. Astrophys. 673, A114 (2023).

23. M. A. Kuhn, L. A. Hillenbrand, A. Sills, E. D. Feigelson,
and K. V. Getman, Astrophys. J. 870, 32 (2019).

24. K. L. Luhman and T. L. Esplin, Astron. J. 160, 44 (2020).

25. K. L. Luhman, Astron. J. 163, 24 (2022).

26. A. Marchal and P. G. Martin, Astrophys. J. 942, 70 (2023).

27. A. M. Mel'nik and A. K. Dambis, Mon. Not. R. Astron.
Soc. 472, 3887 (2017).

28. A.M.Mel'nik and A. K. Dambis, Astron. Rep. 62, 998 (2018).

29. A. M. Mel'nik and A. K. Dambis, Mon. Not. R. Astron. Soc. 493, 2339 (2020).

30. K. F. Ogorodnikov, Dynamics of Stellar Systems
(Fizmatgiz, Moscow, 1965; Pergamon, Oxford, 1965).

31. M. J. Pecaut and E. E. Mamajek, Mon. Not. R. Astron. Soc. 461, 794 (2016).

32. T. Preibish and H. Zinnecker, Astron. J. 117, 2381 (1999).

33. A. Rao, P. Gandhi, C. Knigge, J. A. Paice,
W. C. Leigh, and D. Boubert, Mon. Not. R. Astron. Soc. 495, 1491 (2020).

34. M. J. Sartori, J. R. D. L\'epine, and W. S. Dias, Astron. Astrophys. 404, 913 (2003).

35. L. Thulasidharan, E. D'Onghia, E. Poggio, R. Drimmel,
J. S. Gallagher III, C. Swiggum, R. A. Benjamin,
and J. Alves, Astron. Astrophys. 660, 12 (2022).

36. A. Vallenari, A. G. A. Brown, T. Prusti, et al. (Gaia Collab.), arXiv: 2208.0021 (2022).

37. L. M. Widrow, S. Gardner, B. Yanny, S. Dodelson, and H.-Y. Chen, Astrophys. J. Lett. 750, L41 (2012).

38. N. J. Wright and E. E. Mamajek, Mon. Not. R. Astron. Soc. 476, 381 (2018).

39. N. J. Wright, New Astron. Rev. 90, 101549 (2020).

40. P. T. de Zeeuw, R. Hoogerwerf, J. H. J. de Bruijne,
A. G. A. Brown, and A. Blaauw, Astron. J. 117, 354 (1999).

41. H. Zinnecker and H. W. Yorke, Ann. Rev. Astron. Astrophys. 45, 481 (2007).

 \end{document}